\documentclass[
	reprint,
	superscriptaddress,
	showkeys,
	aps,
	pra,
	longbibliography,
    preprintnumbers, 
	nofootinbib
]{revtex4-2}

\usepackage{subfigure}
\usepackage{mathtools}
\usepackage{amsfonts}
\usepackage{comment}
\usepackage{bbm}
\usepackage{amsmath}
\usepackage{bm}

\usepackage{graphicx}
\usepackage{color}
\usepackage{colortbl}
\usepackage{array}
\usepackage[dvipsnames]{xcolor}

\usepackage{hyperref}
\usepackage[nameinlink]{cleveref}

\usepackage{xifthen}
\usepackage{soul}
\usepackage{xcolor}
\usepackage{enumitem}
\hypersetup{
colorlinks,
linkcolor={red!75!black},
citecolor={blue!75!black},
urlcolor={blue!75!black}
}
\usepackage{units}

\usepackage[utf8]{inputenc}

\graphicspath{{./figures/}}

\newcommand{\gettitle}{Real-time dynamics of false vacuum decay}
\newcommand{\getHeidelbergAffiliation}{\affiliation{Institut f\"ur Theoretische Physik, Universit\"at Heidelberg, Philosophenweg 16, 69120 Heidelberg, Germany}}
\newcommand{\getDesyAffiliation}{\affiliation{Deutsches Elektronen-Synchrotron DESY, Notkestr.\ 85, 22607 Hamburg, Germany}}
\newcommand{\getNorditaAffiliation}{\affiliation{Nordita, Stockholm University and KTH Royal Institute of Technology \\
Hannes Alfvéns v\"ag 12, SE-106 91 Stockholm, Sweden}}

\hypersetup{
colorlinks,
linkcolor={red!75!black},
citecolor={blue!75!black},
urlcolor={blue!75!black},
pdftitle={\gettitle},
pdfauthor={Batini, },
pdfkeywords={XXXX}
bookmarksopen=true,
bookmarksopenlevel=2,
bookmarksnumbered=true
}

\allowdisplaybreaks
\begin{document}
\title{\gettitle}

\author{Laura Batini}
\email{batini@thphys.uni-heidelberg.de}
\getHeidelbergAffiliation

\author{Aleksandr Chatrchyan}
\email{aleksandr.chatrchyan@su.se}
\getDesyAffiliation
\getNorditaAffiliation

\author{Jürgen Berges}
\email{berges@thphys.uni-heidelberg.de}
\getHeidelbergAffiliation

\preprint{DESY-23-158}
\preprint{NORDITA 2023-082}

\begin{abstract}
We investigate false vacuum decay of a relativistic scalar field initialized in the metastable minimum of an asymmetric double-well potential. The transition to the true ground state is a well-defined initial-value problem in real time, which can be formulated in nonequilibrium quantum field theory on a closed time path.  
We employ the nonperturbative framework of the two-particle irreducible (2PI) quantum effective action at next-to-leading order in a large-$N$ expansion. We also compare to classical-statistical field theory simulations on a lattice in the high-temperature regime. By this, we demonstrate that the real-time decay rates are comparable to those obtained from the conventional Euclidean (bounce) approach. In general, we find that the decay rates are time dependent. For a more comprehensive description of the dynamics, we extract a time-dependent effective potential, which becomes convex during the nonequilibrium transition process. By solving the quantum evolution equations for the one- and two-point correlation functions for vacuum initial conditions, we demonstrate that quantum corrections can lead to transitions that are not captured by classical-statistical approximations.

\end{abstract}
\maketitle

\section{Introduction}

False vacuum decay is one of the characteristic phenomena of quantum field theory with a wide range of applications in physics. Examples in particle physics and cosmology include the stability of the current electroweak minimum~\cite{Sher:1988mj, Arnold:1989cb, Elias-Miro:2011sqh, Bezrukov:2012sa,  Markkanen:2018pdo}, the electroweak phase transition~\cite{Anderson:1991zb, Espinosa:1993bs, Farrar:1993hn, Morrissey:2012db}, and inflationary models~\cite{LINDE1982389, Linde:1990flp, Kolb:1990vq, Yamauchi:2011qq}. 
The decay of the false vacuum has been studied in numerous works, building on Langer's theory of bubble nucleation~\cite{PhysRevLett.21.973} and the foundational description by Callan and Coleman~\cite{Coleman:1977py,callan:1977pt,coleman:1985rnk}, which has been extended to thermal field theory by Linde~\cite{Linde:1980tt,linde, Laine:2016hma}. 

In general, the decay of a metastable vacuum consists of a first-order phase transition where the order parameter changes from a metastable phase to a stable phase.
The transition may occur through nucleation of bubbles caused by quantum or statistical fluctuations. Bubbles below a critical size collapse due to the overwhelming surface energy cost, while larger bubbles grow rapidly and eventually fill space to complete the phase transition.

The conventional approach to computing the bubble nucleation rate in field theory involves solving the equations of motion in Euclidean signature to find the saddle point of the path integral (the bounce). While this semiclassical method is expected to work well for strongly first-order phase transitions, when the jump in the order-parameter field at the phase transition is sufficiently large, it has limitations for weaker transitions and breaks down for strongly coupled quantum systems (see, e.g.,~\cite{andreassen_2017,dunne_2005,bergner_2003,bergner_2004}).
Furthermore, this method is limited to describing bubble nucleation and does not capture other processes like spinodal decomposition and resonant nucleation, which can also cause false vacuum decay~\cite{Gleiser:2004iy, goldenfeld2018lectures}.

This work aims to go beyond semiclassical methods for the computation of the decay rate. 
In principle, false vacuum decay is a well-defined initial-value problem in real time, which can be formulated in nonequilibrium quantum field theory on a closed time path~\cite{Keldysh1964DiagramTF}. The problem involves the computation of the time dependence of the order-parameter field from its initial value in the metastable phase to its final value in the stable phase. 

Here, we consider a relativistic scalar quantum field theory with quartic self-interaction in 3+1 space-time dimensions. We compute the time-dependence of the field using the two-particle irreducible (2PI) effective action at next-to-leading order (NLO) based on a nonperturbative large-$N$ expansion~\cite{berges_2002,aarts_2002, arrizabalaga_2004}. In the high-temperature regime, we also compute the real-time dynamics from an approximation based on classical-statistical field theory simulations on a lattice~\cite{Braden:2018tky, Pirvu:2021roq, Tranberg:2022noe}.

This allows us to establish a ``dictionary'' between semiclassical characterizations of false vacuum decay and nonequilibrium field theoretical descriptions in terms of time-dependent order-parameter fields and correlation functions. Importantly, to formulate a quantitative comparison of the decay rate, we properly take into account fluctuations using an effective potential, building upon prior work in this direction~\cite{PhysRevD.62.085013, Braden:2022odm,Alford:1993ph,Laine:2016hma,Gould:2021ccf}.

This work aims to identify relevant and measurable information on the decay that can be extracted and constructed from correlation functions, which is particularly practical for situations in which semiclassical concepts, such as the identification and counting of bubbles, become ambiguous.
This is especially relevant given the increased experimental possibilities of tabletop experiments. These systems offer a very high degree of controllability and a genuinely quantum-mechanical nature, making them unique platforms for analogue simulations of quantum field dynamics~\cite{Zache:2019xkx, Prufer:2019kak}, including the decay of a false vacuum, see, e.g., experimental proposals for false vacuum decay using ultracold atoms~\cite{fialko_2015, fialko_2017, Braden:2017add, Billam:2018pvp,Billam:2021nbc, Billam:2022ykl, Jenkins:2023pie}, quantum spin chains~\cite{Lagnese:2021grb, Lagnese:2023xjg}, and a recent experiment using a ferromagnetic superfluid \cite{Zenesini:2023afv}. 

This paper is organized as follows. After introducing the false vacuum decay problem in Sec.~\ref{Problem}, in Sec.~\ref{Thermal}, we summarize relevant aspects of the Euclidean framework to describe phase transitions.
Sec.~\ref{FalsevacuumdecayT} considers the dynamics for a high-temperature situation, where classical-statistical fluctuations are expected to dominate over quantum fluctuations. In Sec.~\ref{Decay_rate}, we establish a link between Euclidean and real-time decay rates in a system at high temperatures.
Section~\ref{sec:Quantum} addresses the problem of real-time dynamics in the quantum regime at low temperatures, where we discuss the importance of genuine quantum corrections to the real-time dynamics using the 2PI effective action approach. Finally, we conclude and give an outlook in Sec.~\ref{sec:Conclusion}.
Several appendices provide details on quantum field theory out of equilibrium and the derivation of the time evolution equations.

\section{The problem of false vacuum decay in field theory} \label{Problem} 
\begin{figure}[t]
\includegraphics[width=.8\linewidth]{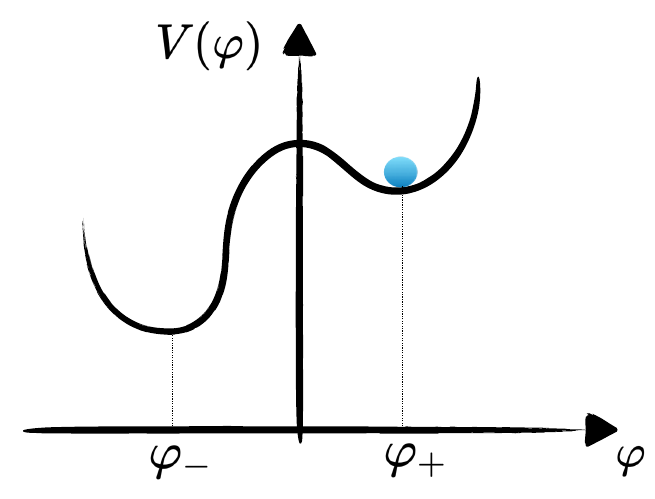}
\caption{Sketch of the classical potential $V(\varphi)$ in \labelcref{Bare}.}
\label{fig:eff potential}
\end{figure}

The system we consider is a (3+1)-dimensional scalar field theory with classical action\footnote{Throughout this work we use units where $\hbar = c = k_B = 1$.}
\begin{equation}
\label{eq:classact}
S[\varphi]=\int \mathrm{d}^4  x\left\{\frac{1}{2}\dot \varphi^2-\frac{1}{2}(\nabla \varphi)^2-V(\varphi)\right\} \, ,
\end{equation}
for a scalar field $\varphi(x^0,\boldsymbol{x})$ depending on time $x^0 \equiv t$ and spatial coordinates $\boldsymbol{x}$, whose potential is given by
\begin{equation}
V(\varphi)=- \frac{1}{2} m^2\varphi^2+\frac{\lambda}{4!} \varphi^4 + h \varphi \, . 
\label{Bare}
\end{equation}
The parameters $m^2, \lambda, h$ are real and positive. The potential with $h=0$ 
exhibits two degenerate minima at constant values of $ \varphi$ given by $\varphi_{\pm}(h=0) = \pm \sqrt{6 m^2/\lambda}$ separated by a potential barrier. 
We consider $h\neq 0$ to lift this degeneracy, such that a double-well potential arises with two minima whose free-energy difference is nonzero, $V( \varphi_+(h \neq 0)) - V( \varphi_-(h \neq 0)) \neq 0$, as sketched in~Fig.~\ref{fig:eff potential}. The local minimum $\varphi_+$ denotes the “false vacuum” in a first-order phase transition to the global minimum, or “true vacuum,” $\varphi_-$.

False vacuum decay is an initial-value problem, starting from a nonequilibrium initial condition (the metastable state).
If one only considers classical dynamics, the time evolution of the field is governed by the classical equation of motion, i.e.,
\begin{equation}
    \left[\partial_t^2-m^2+\frac{\lambda}{6} \varphi^2(t)\right] \varphi(t) +h =0 \, ,
    \label{eq:classvarphi}
\end{equation}
if spatial homogeneity is assumed.
If the initial field value is close to the local minimum $\varphi_+$, then the field may not reach the global minimum without additional kinetic energy to overcome the potential barrier. 
In quantum field theory, however, the corresponding state is always expected to decay towards the state with the lowest available free energy. In general, the description of the transition requires the consideration of quantum and/or statistical fluctuations. 
 Therefore, both the initial value $\varphi_+$, as well as the dynamical evolution Eq.~\labelcref{eq:classvarphi}, and thus the final field value $\varphi_-$, will receive corrections due to quantum-statistical fluctuations.
 At very low temperatures or vacuum initial conditions, the transition is dominated by quantum fluctuations, where tunneling plays an essential role.

In quantum-statistical field theory, the initial pure or mixed state of the system can be described by a density operator $\varrho(t_0)$ at a given time $t_0$. The goal is to understand the temporal evolution of the system from its initial state at $t_0$ to its subsequent states at later times. The expectation value of any observable $\mathcal{O}(\varphi)$ and its dynamical evolution can be derived from the functional integral~\cite{calzetta_hu_2008}
\begin{equation}
\label{eq:expval}
\langle\mathcal{O}(\varphi)\rangle=\int \mathcal{D} \varphi \varrho\left(t_0\right) \mathcal{O}(\varphi) e^{i S[\varphi]} \, ,
\end{equation}
whose definition involves the Schwinger-Keldysh closed time path~\cite{Keldysh1964DiagramTF}, see Appendix~\ref{NEQQFT} for more details. 
For instance, for $\mathcal{O}(\varphi) = \varphi $ the path integral \labelcref{eq:expval} would encode the evolution equation for the field expectation value $\phi(t) = \langle \varphi \rangle(t)$ which, for a spatially homogeneous system, now involves the quantum-statistical fluctuations in the initial state and the corrections to the classical evolution Eq.~\labelcref{eq:classvarphi}. The time-dependent expectation value $\phi(t)$ evolves during the transition from its initial value $\phi_+ = \phi(t_0)$ to its final value $\phi_- = \phi(t)$ at sufficiently later times $t > t_0$.

In general, the functional integral \labelcref{eq:expval} is too complicated to be solved numerically without further approximations. Because of the rapidly oscillating integrand $\sim$$e^{i S}$, standard numerical techniques based on importance sampling cannot be applied, and one has to resort to suitable approximate descriptions, which we turn to in the following sections.
\section{Conventional Euclidean approach to false vacuum decay} \label{Thermal} 

 This section summarizes the most important calculation steps within the conventional Euclidean approach to false vacuum decay. This is to prepare for the comparison with the real-time analysis in the following sections.
 The central observable here is the decay rate per unit volume. The latter consists of the probability of nucleation of bubbles of the new phase per unit volume per unit time.

\subsection{Bounce}
Let us consider a theory defined by the classical action \labelcref{eq:classact}.
The Euclidean action is obtained by analytically continuing time to imaginary values by performing the Wick rotation $\tau = i t$. It is written as $S_E = - i S$ and is a functional of the Euclidean field $\varphi_E(\tau,\boldsymbol{x})$: 
\begin{equation}
    S_E=\int \mathrm{d} \tau \mathrm{d}^3 x \left\{\frac{1}{2}\left(\frac{\mathrm{d} \varphi_E}{\mathrm{d} \tau}\right)^2+ \frac{1}{2}(\nabla \varphi_E)^2 + V(\varphi_E)\right\}
    \, .
\end{equation}
In the Euclidean action, the sign preceding the potential term is reversed compared to the Minkowskian action. Due to this ``inverted potential,'' the Euclidean action allows for nontrivial classical solutions connecting the two minima. One of these solutions is crucial in describing false vacuum decay, as described in the following.

To compute the decay rate per unit volume, one can expand the Euclidean action around the saddle-point solution $\overline{\varphi}_B(\tau,\boldsymbol{x})$, which satisfies the classical Euclidean field equation of motion 
 \begin{equation}
 \frac{\delta S_E}{\delta \varphi_E}\Big|_{\varphi_E=\overline{\varphi}_B}=0
 \, ,
 \end{equation}
 together with the boundary conditions
\begin{equation}
   \quad \overline{\varphi}_B(0, \boldsymbol{x})=\overline{\varphi}_B(\beta, \boldsymbol{x}), \quad \lim _{|\boldsymbol{x}| \rightarrow \infty} \overline{\varphi}_B(\tau\, , \boldsymbol{x})=\varphi_+\, ,
   \label{eq:Stationarity}
\end{equation}
where $\beta = 1/T$ is the inverse temperature. 

The field configuration $\overline{\varphi}_B$, often called bounce or instanton, is associated with the critical bubble within the metastable phase, the interior of which consists of the stable phase.
We rewrite the stationarity condition \labelcref{eq:Stationarity} by assuming four-dimensional rotational symmetry of the solution in the limit of vanishing temperature. 
Thus, the bounce equation in terms of the hyperradial coordinate $\rho^2 = x^2 +y^2 +z^2+\tau^2$ simplifies to the one-dimensional differential equation
   \begin{equation}
   \label{eq:4d}
    \frac{d^2 \varphi_E}{d \rho^2}+\frac{3}{\rho} \frac{d \varphi_E}{d \rho}=V^{\prime}\left(\varphi_E\right) \, ,
\end{equation}
where prime denotes $d/d \varphi_E$.
This equation describes the motion of a particle in the inverted potential and subject to a “friction force” $\frac{3}{\rho} \frac{d \varphi_E}{d \rho}$.

On the other hand, in the high-temperature limit, the leading contribution may be obtained from the dimensionally reduced theory by performing the integration over the Euclidean time coordinate, i.e.,
\begin{equation}
    S_E[{\varphi_E}]=
   \frac{S_{3}[{\varphi_E}]}{T}
   \, .
\end{equation}
Here, $S_{3}[{\varphi_E}]$ denotes the rescaled action of a three-dimensional field theory for a $\tau$-independent field. In this case, the critical bubble configuration has a three-dimensional rotational symmetry. It can be parametrized by $\overline{\varphi}_B(r)$, with $r^2 = x^2 +y^2 +z^2$, the solution of the bounce equation similar to the previous case:
\begin{equation}
\label{eq:3d}
    \frac{d^2 \varphi_E}{d r^2}+\frac{2}{r} \frac{d \varphi_E}{d r}=V^{\prime}\left(\varphi_E\right)
    \, .
\end{equation}
 In both the high-$T$ and low-$T$ cases, the bounce solutions satisfy the boundary conditions  
\begin{equation}
    \overline{\varphi}_B \rightarrow \varphi_+ \text{   \quad for  \quad  } r, \rho \rightarrow \infty \, ,
\end{equation}
and 
\begin{equation}
\overline{\varphi}_B^{\prime}=0 \text{   \quad for  \quad   } r, \rho =0 
\, ,
\end{equation}
 for the respective radial coordinate. 
 The corresponding explicit expression for the high-$T$ Euclidean action in polar coordinates is given by
\begin{equation}
 \overline{S}_3 = S_3[\overline{\varphi}_B] = 4 \pi \int_0^{\infty}  \mathrm{d} r r^2 \left\{ \frac{1}{2}\left(\partial_r \overline{\varphi}_B\right)^2+V( \overline{\varphi}_B)\right\} 
 \, .
 \label{eq:bounce}
\end{equation}

 \subsection{Nucleation rate}

The knowledge of the bounce action allows one to estimate the decay rate. We now consider fluctuations around the non-trivial saddle point, described by \labelcref{eq:4d} and \labelcref{eq:3d}, respectively.
The actual calculation of the one-loop fluctuation determinant involves one negative eigenmode, several zero modes and infinitely many positive modes. Restricting to the limits of low- and high-temperature cases, the nucleation rate per unit volume for the former case can be written as~\cite{Coleman:1977py}
 \begin{equation}
     \left.\frac{\overline \Gamma}{V_3}\right|_{\text {low-} T} \simeq\left(\frac{\overline{S}_E}{2 \pi }\right)^{2}\left|\frac{\operatorname{det}^{\prime}\left[-\partial^2+V^{\prime \prime}(\overline{\varphi}_B)\right]}{\operatorname{det}\left[-\partial^2+V^{\prime \prime}(\varphi_+)\right]}\right|^{-\frac{1}{2}} e^{-\overline{S}_E}.
     \label{Gamma_over_V_formula}
 \end{equation}
Here, $\overline{S}_E \equiv S_E[\overline{\varphi}_B]-S_E[\varphi_+]$. The prime on the determinant indicates that the zero eigenvalues, i.e., the translational modes, have been integrated out (but the negative eigenmode is kept)~\cite{Laine:2016hma}. In this case, due to the symmetry of the solution, there are four zero modes, each bringing a factor of $(\frac{\overline{S}_E}{2 \pi})^{\frac{1}{2}}$ and resulting in the factor $(\frac{\overline{S}_E}{2 \pi})^2$ in \labelcref{Gamma_over_V_formula}. Note that the decay rate per unit volume is exponentially suppressed by $\overline{S}_E$.

In the high-$T$ case, the decay rate per unit volume is expected to be given by~\cite{Linde:1980tt}

\begin{equation}
\left.\frac{\overline \Gamma}{V_3}\right|_{\text {high-} T}\!\!\!\simeq\frac{|E_-|}{2 \pi}\!\left(\frac{\overline{S}_3}{2 \pi T}\right)^{\frac{3}{2}}\left|\frac{\operatorname{det}^{\prime}\left[-\partial^2+V^{\prime \prime}(\overline{\varphi}_B)\right]}{\operatorname{det}\left[-\partial^2+V^{\prime \prime}(\varphi_+)\right]}\right|^{-\frac{1}{ 2}} \!\!\! e^{-\frac{\overline{S}_3}{T}},
\label{Eq_rate_highT}
\end{equation}
where $\overline{S}_3 \equiv S_3[\overline{\varphi}_B]-S_3[\varphi_+]$ and $E_-^2$ indicates the single negative eigenvalue. The different exponent $3/2$ in the prefactor compared to the low-$T$ case is due to the three translational modes in the high-temperature regime.

In typical cases, the action appearing in the exponents in \labelcref{Gamma_over_V_formula} or \labelcref{Eq_rate_highT} is significant, such that the exponential is very small. Therefore, the order of magnitude of $\overline{\Gamma}/V_3$ is determined predominantly by the instanton action and the temperature. In this work, we concentrate on this exponential part and disregard the prefactor, which may be expected to be of the order of unity.

In order to make practical use of the above formulae, it is necessary to calculate the bounce. However, relying solely on a saddle point approximation around the bounce obtained from the classical potential $V$ may not provide an accurate description. 
The impact of finite temperature can be significant because thermal fluctuations modify the potential, leading to a finite-temperature effective potential. In other cases, the classical action may not exhibit two distinct minima, but quantum corrections introduce an additional local minimum in the effective potential~\cite{Strumia:1998vd, Moore:2000jw}. Therefore, incorporating the quantum-statistical fluctuations into a modified effective potential can offer a more efficient description~\cite{Alford:1993ph, GLEISER1993271, Berges:2000ew, Braden:2022odm, Laine:2016hma, Gould:2021ccf}. 

For the underlying real-time problem of false vacuum decay, due to its nonequilibrium nature, it can also be necessary to consider the effective potential as time dependent. As a consequence, the bounce and the decay rate per unit volume can, in principle, vary with time. Therefore, a single decay rate may not be enough to fully capture the nonequilibrium process's complexity. We consider in the next section the nonequilibrium real-time problem in a high-temperature scenario to address these questions.

\section{Real-time false vacuum decay in the classical-statistical regime} \label{FalsevacuumdecayT}
\begin{figure*}[t!]
\centering
\includegraphics[width=.75\textwidth]{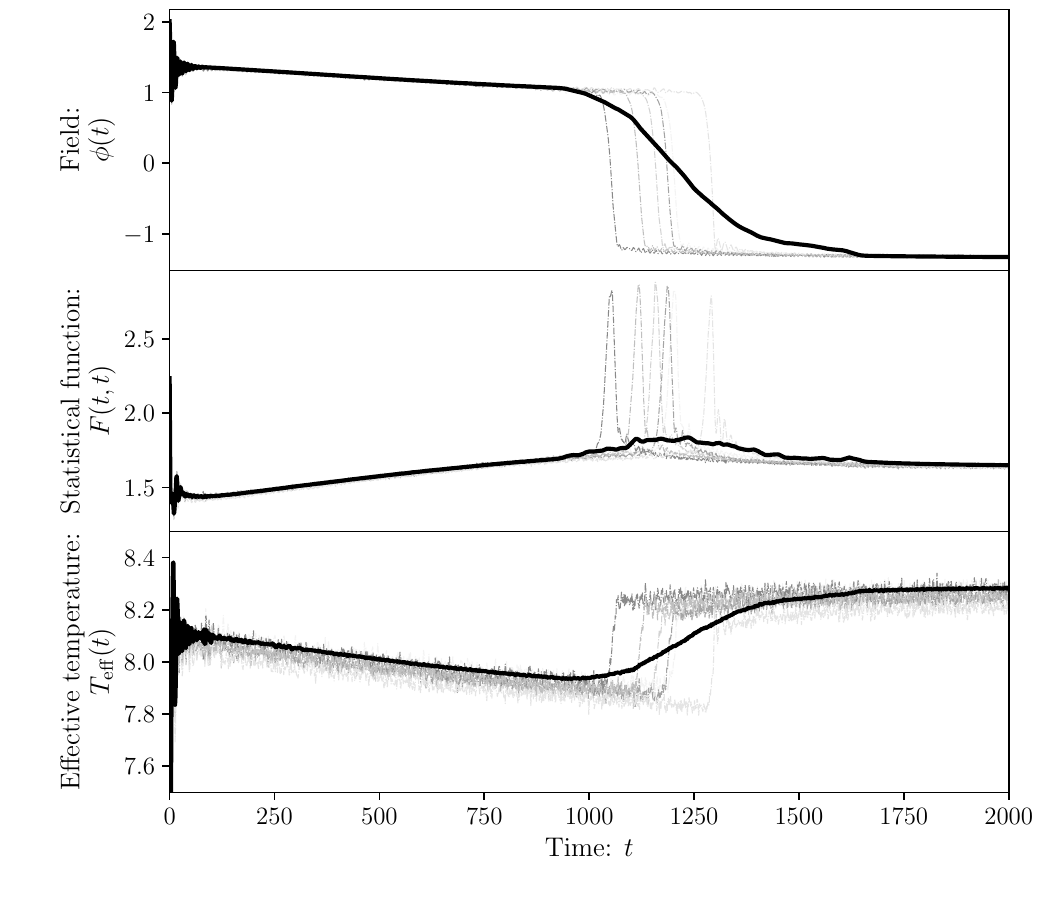}
\caption{Time evolution of the one-point function $\phi(t)$, the equal-time statistical propagator $F(t,t)$, and the effective temperature $T_{\text{eff}}(t)$. The black solid lines represent the correlation functions obtained from averages over $n=200$ runs, while the gray dashed lines show a subset of single realization runs.}
\label{fig:CS n-point functions}
\end{figure*}
In this section, we analyze the evolution in time of a field theory characterized by the classical action \labelcref{eq:classact}. Initially, the field is trapped in the false minimum of the potential \labelcref{Bare}. We assume that the system is prepared in an initial state at a high temperature, represented by the density operator $\varrho(t_0)$. This choice allows us to approximate
the dynamics through classical-statistical field theory evolution~\cite{berges2015nonequilibrium}.

The idea behind the classical-statistical field description is that the field is evolved according to the classical real-time equations of motion. The quantum expectation values are then obtained by averaging over the statistical ensemble with the initial correlation functions determined by the quantum theory.

\subsection{Dynamics of the field}
We introduce dimensionless quantities
\begin{equation}
  t \rightarrow t m \, , \,\, x \rightarrow x m \, , \,\,  \varphi \to \frac{m}{\sqrt{\lambda}} \varphi \, , \,\,  h \to \frac{m^3}{\sqrt{\lambda}} h \, ,
\end{equation}
such that the parameters in the potential can be reduced to only one dimensionless parameter by rescaling as 
\begin{equation}
 V \rightarrow \frac{m^4}{\lambda} V  = -\frac{1}{2}  \varphi^2+\frac{1}{4!} \varphi^4 + h \varphi \, .
\end{equation} 
The dimensionless parameter $h$ indicates how much the potential is tilted.
We prepare the system's initial state with a spatially homogeneous field average close to the potential's local minimum. More precisely, we set for the field at initial time $\phi(t=0)=\phi_0$, $\dot{\phi}(t=0)=0$. The initial fluctuations are encoded via the initial two-point correlation function. These fluctuations change the effective potential of~\labelcref{Bare} and, as a result, the field is typically displaced from the local minimum of the effective potential at initial time.

The subsequent time evolution of the system can be divided into several phases. 
It typically takes some time for the transition to the global minimum of the effective potential to begin.
During the early-time dynamics of the field average or one-point function, the system oscillates around the local minimum of the effective potential. As the energy is converted into fluctuations with nonzero momentum, these oscillations gradually decrease in amplitude. During this period, a resonating momentum band gets excited and at later times, the oscillations decay.
In the second, much longer phase, the system evolves towards a prethermal state with a field value close to the local minimum. Prethermal means that the system has many equilibrium properties but is not yet  thermalized~\cite{Berges:2004ce}, as the field is still trapped in the false minimum. In particular, the field obeys the equipartition of energy between momentum modes, as expected for a classical system. 
This prethermalization has been verified in our simulations, where we observed that the average kinetic energy of each mode is ${T_{\mathrm{eff}}}/{2}$.
This allows us to define an effective temperature $T_{\text{eff}}$ of the system as
\begin{equation}
    \frac{T_{\text{eff}}(t)}{2} \equiv  \frac{1}{2} \langle{{\dot 
    		\varphi(t, \boldsymbol{x})^2} \rangle}  \, ,
    \label{eq:effective temperature}
\end{equation}
where the brackets indicate the ensemble average and the average over volume is implied.
In this second phase, the system effectively loses the details of the initial conditions. At the same time, the dynamics depends on only a few parameters: the couplings of the effective potential, which differ from the bare potential due to fluctuations, and the total energy density.
Consequently, the specific initialization of the fluctuations in the system becomes less important at later times.

For $h\neq 0$, the field eventually transitions from the local to the global minimum. The timescale at which the transition takes place depends on the value of $h$. In particular, for large values of $h$, the transition can occur before establishing prethermalization. In contrast, for small values of $h$, the transition occurs when the system is already nearly thermal. The strength of the fluctuations or, equivalently, the effective temperature is another parameter determining the transition time.

Numerical results are obtained from simulations on a cubic lattice with periodic boundary conditions. The number of lattice sites per spatial dimension and the spacing are denoted as $N_x$ and $a_x$, respectively,  such that the linear lattice size is $L=N_x a_x$. We set $a_t = 0.01 a_x$, $N_x=64$,  $h = 0.055$. The initial conditions for the field were chosen as
 \begin{equation}
 	\varphi(0, \boldsymbol{x}) = \phi_0 + \int \frac{d^3 p}{(2 \pi)^3} \sqrt{\frac{ f_{\boldsymbol{p}}(0)}{\omega_{\boldsymbol{p}}(0)} } c_{\boldsymbol{p}} e^{i \boldsymbol{p x}} \, ,
 \end{equation}
where $\phi_0 \equiv \langle \varphi(0, \boldsymbol{x})\rangle= 1.99$. We choose the initial distribution as $f_{\boldsymbol{p}}(0) = 2.225 \, \omega_{\boldsymbol{p}}(0)$, and $\omega_{\boldsymbol{p}}(0)= \sqrt{ \boldsymbol{p}^2 +m_{init}^2}$ is the initial frequency and $c_{\boldsymbol{p}}$ satisfies $\langle c_{\boldsymbol{p}} c^*_{\boldsymbol{p^{\prime}}} \rangle= (2\pi)^3 \delta(\boldsymbol{p} - \boldsymbol{p}^{\prime} ) $~\cite{berges2015nonequilibrium}. 

For these values, the time evolution of $\phi(t)$, the equal-time statistical connected two-point function, defined as $F(t,t) \equiv   \int_{\boldsymbol{p}} F(t,t, \boldsymbol{p}) =  \langle \varphi(t, \boldsymbol{x})^2 \rangle   -   \phi(t)^2$, and the effective temperature $T_{\text{eff}}(t)$ is shown in Fig.~\ref{fig:CS n-point functions}.
The black solid lines show the correlation functions obtained from averages over $n= 200$ runs. The gray dashed lines indicate a subset of single realization runs, which are not averaged, for comparison. One observes a short phase of decaying oscillations of $\phi(t)$ at very early times. Next, in the more prolonged phase before the transition, the field value decreases almost linearly, together with the temperature. We emphasize that this stage of the dynamics depends on the choice of the initial conditions. The white noise initial spectrum, considered here, results in energy flow in momentum space towards the infrared region. This leads to the observed slow growth of $F(t,t)$, which, in turn, decreases the value of $\phi(t)$. Simultaneously, the temperature $T_{\text{eff}}$ time slightly decreases. Eventually, the transition to the true minimum begins, resulting in the faster decrease of $\phi(t)$.

\begin{figure}[t]
  \centering
  \begin{minipage}{4cm}
    \centering
    \includegraphics[width=4cm]{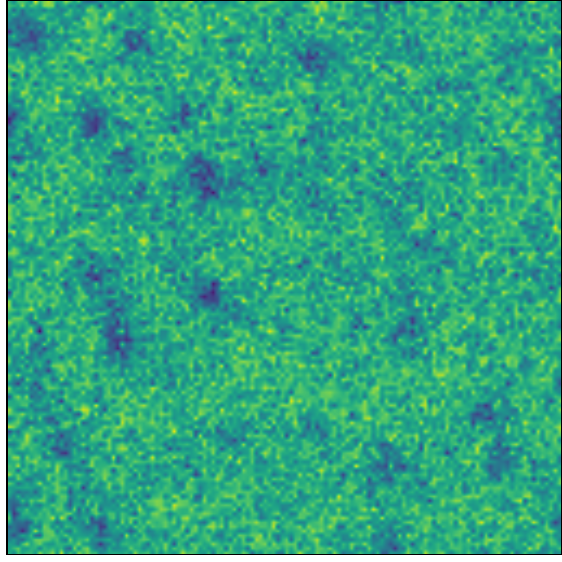}
  \end{minipage}
  \begin{minipage}{4cm}
    \centering
    \includegraphics[width=4cm]{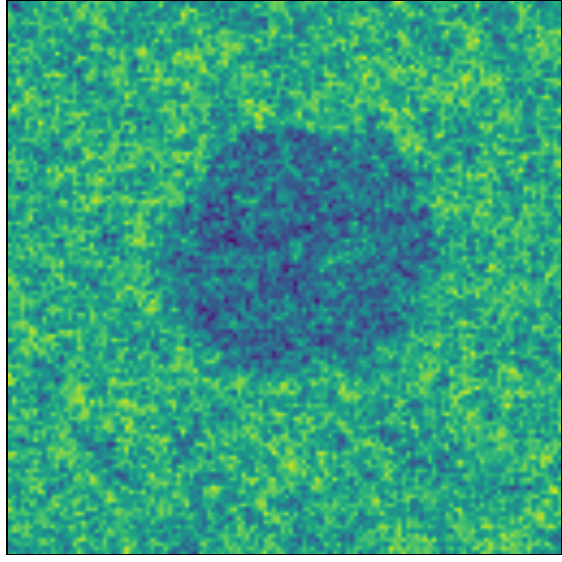}
  \end{minipage}
  \begin{minipage}{4cm}
    \centering
    \includegraphics[width=4cm]{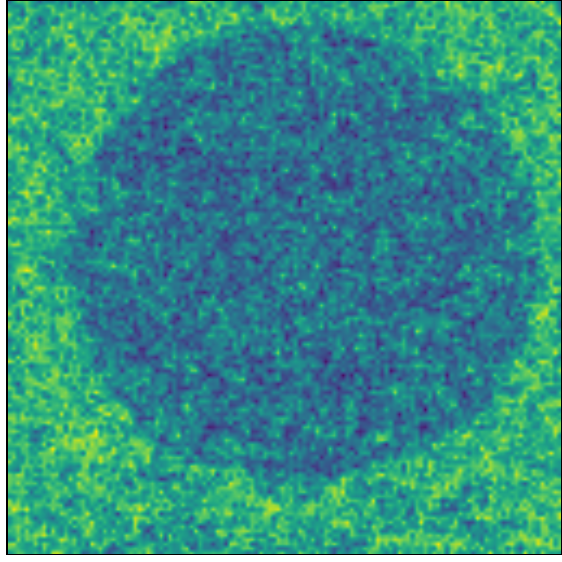}
  \end{minipage}
  \begin{minipage}{4cm}
    \centering
    \includegraphics[width=4cm]{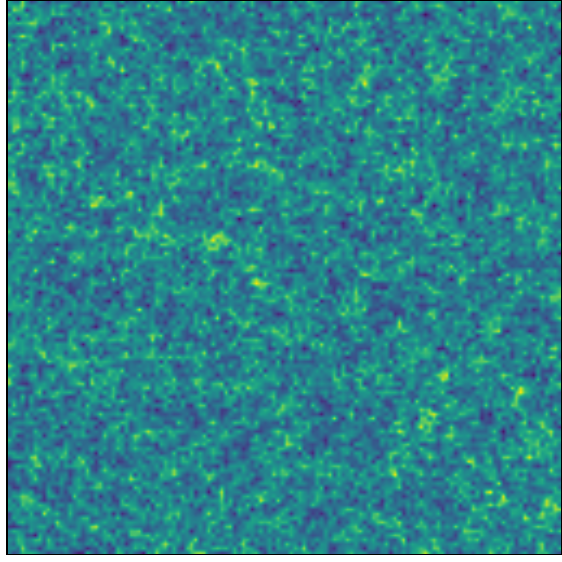}
  \end{minipage}
  \caption{Snapshots of the evolution for the classical field $\varphi(t,\boldsymbol{x})$ from a two-dimensional slice using a single realization run: (top left) metastable phase, (top right) bubble nucleation, (bottom left) bubble expansion, and (bottom right) stable phase.}
  \label{fig:snapshots}
\end{figure}

The decay of the false vacuum can proceed via bubble nucleation in position space. The formation of bubbles can be observed in the evolution of individual classical-statistical simulation runs. To illustrate this, snapshots of the evolution for a two-dimensional section of the classical field are shown in Fig.~\ref{fig:snapshots} using a single run without averaging. At the beginning (top left), the background is filled with the metastable phase (lighter shade) in the presence of fluctuations (darker shade). 
Finally, in the last portion of the trajectory, the field oscillates around the true ground state and starts thermalizing (bottom right). 
In the snapshots of the real-space configurations, a bubble of the new phase exceeding the critical size appears (top right) and starts expanding (bottom left). 
In the following subsection, we focus on how to extract the decay rate using the notion of an effective potential.  

\subsection{Time-dependent effective potential} \label{subsec:Effective parameters}

To calculate the exponential factor that governs the decay rate per unit volume $\overline{\Gamma}/V_3$ 
 in~\labelcref{Eq_rate_highT}, one has to compute the effective temperature for the system $T_{\text{eff}}$ and the shape of the bounce $\overline{\varphi}_B$, which in turn depends on the (effective parameters of the) potential $V(\varphi)$. 
As described in the following, we consider a method for extracting the effective potential directly from the lattice simulations.
In a regime with slow field evolution, such as realized in the prethermal regime described above, kinetic energy contributions should be small, i.e., $\dot \phi \ll m \phi$ and $\ddot \phi \ll m^2 \phi$. We may extract an effective free energy in terms of the macroscopic field $\phi$. 

More precisely, up to the time when the first bubble is nucleated, we consider a free energy or effective potential $V_{\text{eff}}(\phi)$ of the same form as the bare one, but with effective couplings: 
\begin{equation}
V_{\text{eff}}(\phi) = -\frac{1}{2}m_{\text{eff}}^2\phi^2 + \frac{1}{4 !}\lambda_{\text{eff}} \phi^4  + J \phi \, . 
\label{Derivativeeff}
\end{equation}
It involves the effective mass $m_{\text{eff}}$ and effective coupling $\lambda_{\text{eff}}$. We also allow for a constant source term $\sim$$J$ linear in the field to probe the effective potential at different values of $\phi$. 
Specifically, for each external source $J$, we denote the value the field settles at a given time $t$ in $\phi = \phi(J) = \langle \varphi(J) \rangle$. This field value corresponds to the stationary point of the effective potential, where its derivative with respect to $\phi$ vanishes, i.e., $\partial V_{\text{eff}}(\phi)/\partial \phi = 0$. Inversion reveals a one-to-one correspondence between each value of $\phi$ and the corresponding linear term, denoted as $J = J(\phi)$, explicitly given by
\begin{equation}
  J(\phi) =  m^2_{\text{eff}}  \phi  - \frac{1}{6} \lambda_{\text{eff}} \phi  ^3  \, .
\label{eq:derivative_potential_effective}
\end{equation}
Our analysis for extracting the effective potential is similar to the one in~\cite{PhysRevD.62.085013}, where the time-evolution projection of the microscopic equation of motion onto the zero momentum sector is considered.
In particular, they consider the volume-averaged field value in a single run. As a consequence, the field is subject to a stochastic force in addition to the deterministic one. Our analysis, instead, is based on tracking the ensemble average field rather than a single run averaged over the volume. As a result, the stochastic component vanishes, leaving only the deterministic parts. Also, the second time derivative of the field becomes significantly smaller because the oscillations average out within the ensemble.

\begin{figure}[t]
\centering
\includegraphics[width=0.99\linewidth]{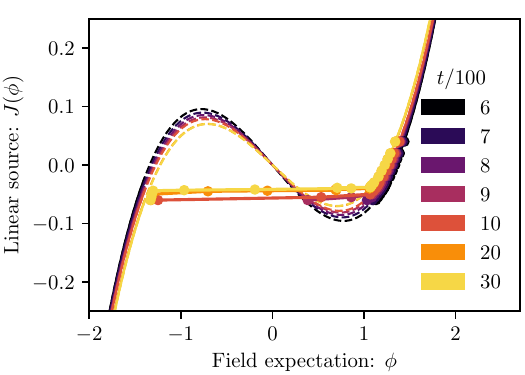}
\caption{Linear source $J$ as a function of the field expectation value $\phi$ at various times $t$. The solid lines represent the fitted results obtained in the prethermal state, i.e., before decay occurs, using \labelcref{eq:derivative_potential_effective}. The data points in the inner region of the potential deviate from the fit, as indicated by the dashed lines. This observation reveals a convex shape for the inner part of the potential.}
\label{fig:Derivativepot}
\end{figure}

The linear term as a function of the field expectation value is illustrated in Fig.~\ref{fig:Derivativepot} using circles for various times, before and after the transition.

Using \labelcref{eq:derivative_potential_effective}, we can thus estimate the effective couplings $m^2_{\text{eff}}$ and $\lambda_{\text{eff}}$ from our numerical simulations. 
The effective parameters are determined by minimizing the sum of the squares of the residuals. The solid lines depicted in  Fig.~\ref{fig:Derivativepot} represent the fitted values of the derivative of the effective potential. We observe that the effective couplings exhibit a slow evolution over time.

While the extracted data points in Fig.~\ref{fig:Derivativepot} appear to be well described by \labelcref{Derivativeeff} in the outer convex regions of the potential, this is not the case in the inner nonconvex part of the polynomial ansatz. In the inner region, indicated by the dashed lines in Fig.~\ref{fig:Derivativepot}, the convexity of the effective potential implies a nonanalytic form beyond \labelcref{Derivativeeff}. Accordingly, the data points extracted from the simulations deviate from the fit in the inner region. 
The coexistence of two hysteresis phases suggests the presence of this inner region and is typically associated with the notion of ``Maxwell's construction.'' 

\section{Comparison of the real-time decay with the Euclidean approach} \label{Decay_rate}
This section compares the real-time evolution of false vacuum decay explored in the previous section and the Euclidean approach discussed in Sec.~\ref{Thermal}. We focus on the decay rate per unit volume, the fundamental quantity for our analysis. In the Euclidean approach, the rate per unit volume for the specific high-temperature scenario can be approximated as $\overline{\Gamma}/V_3\sim e^{-\overline{S}_3/T_{\mathrm{eff}}}$, where $\overline{S}_3$ represents the three-dimensional bounce action, and $T_{\mathrm{eff}}$ is the effective temperature.

  Before analyzing the numerical data, it is essential to understand how the decay rate is affected by the timescales of the system: the prethermalization timescale (the timescale for the prethermalization processes in the metastable phase to complete) and the transition timescale (the timescale for the nucleation of a bubble to occur). Our central question focuses on the possibility of capturing the true real-time dynamics through a single rate and how to extract it from the real-time data.
  In the case of prethermalization times significantly longer than the transition timescale, the decay rate will be independent of time.
  However, for practical, computational or experimental observations, it is essential to choose parameter values that allow us to observe the transition at a reasonable time. As a result of this choice, the two timescales may not be so different, leading to some time dependence of the decay rate. 

\subsection{Extracting the decay rate from the effective potential}
\begin{figure}[t]
\centering
\includegraphics[width=0.99\linewidth]{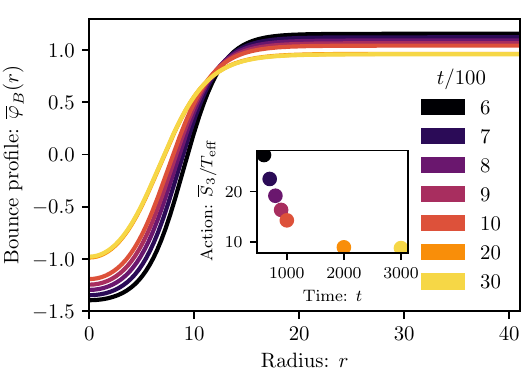}
\caption{Profiles of the bounce $\overline{\varphi}_B(r)$ as a function of the radius $r$ are shown for various selected times $t$, as presented in Fig.~\ref{fig:Derivativepot}. The inset displays the corresponding ratio between the bounce action $\overline{S}_3$ and the effective temperature $T_{\text{eff}}$, obtained using \labelcref{eq:effective temperature}.}
\label{fig:bounce}
\end{figure}
 To extract the relevant timescales for computing the rates, we identify the onset of the decay process at approximately $t_0 \approx 800$. This is evident from the data points deviating from the polynomial fit, as shown in  Fig.~\ref{fig:Derivativepot}. Our primary focus is calculating the rate per unit volume around this time $t_0$, including a small time interval before and after. Note that at much later times the field reaches the true minimum, and the decay process is already complete, rendering the concept of a decay rate irrelevant. Conversely, the decay rate is expected to be extremely small for much earlier times before the decay starts. We will investigate this point in the following subsections.

Following Sec.~\ref{Thermal}, we determine the profiles of the critical bubble $\overline{\varphi}_B(r)$ for various times. Rarely can \labelcref{eq:3d} be solved analytically. Thus, we numerically compute the associated bounce using the so-called shooting method~\cite{Coleman:1977py,KONG2021399}. The solutions are shown in Fig.~\ref{fig:bounce}, where a clear time dependence of the bounce profile is observed. For each time, we calculate the corresponding bounce action $\overline{S}_3(t)$ by inserting the profile into \labelcref{eq:bounce}, and the time-dependent effective temperature $T_{\text{eff}}(t)$ is determined using \labelcref{eq:effective temperature}. The resulting ratio $\overline{S}_3(t) / T_{\text{eff}}(t)$ is displayed in the inset of Fig.~\ref{fig:bounce}. Notably, this ratio decreases with time and only converges to a time-independent value at significantly later times.

The corresponding decay rate, defined as $(\Gamma/V_3)^{1/4}$, has the units of inverse time.
The result is illustrated in Fig.~\ref{fig:methodscomparison}  as the gray solid line. 
  This decay rate is depicted for the aforementioned time range and will be used for later comparison.
  We note again that this result does not consider any corrections to the rate that come from the preexponential factor.

\subsection{Extracting decay rates from real-time dynamics}
We have developed two methods to compute the decay rate from our real-time simulations. 

\subsubsection{Survival probability}
The first method we employ to find the decay rate involves monitoring the formation of bubbles as they nucleate in specific regions of space. To identify the nucleation of bubbles, we conduct a series of classical-statistical simulations on cubic lattices of various volumes $V_3=128^3, 256^3, 512^3$. We maintain identical parameters in each simulation and initiate the system in the metastable state. We then observe the elapsed time until the system grows a bubble of the stable phase, denoting it the nucleation time $t$.
To ensure the statistical significance of our results, we perform a total of $n=125$ simulations for each volume.
To resolve a critical bubble, we consider a small lattice cube and compute the average field value inside it for each point on the lattice. The size of the cubes is chosen according to the expected radius of the bubbles, estimated to be around $r \sim 10$ (see Fig.~\ref{fig:bounce}). The primary purpose is to prevent the doublecounting of the same bubble during the nucleation process. While the specific length of the cubic boxes is not critical, it is chosen to ensure that the results are independent of this scale. The system is considered to have decayed when one (or more) of these cubic boxes exhibit an average field value greater than the value $\phi_{\text{max}} \approx 0$. 

We create a time distribution based on collected data of bubble nucleation events and then calculate the cumulative distribution, also known as survival probability. 
The survival probability is defined as $p(t) \equiv n_{\text{surv}}(t)/n$, where $n_{\text{surv}}(t)$ represents the number of simulations that have not nucleated an expanding bubble up to a given time $t$.

By utilizing the cumulative distribution instead of the decay time distribution, as done in~\cite{Borsanyi:2005mb}, we gain an advantage in studying the distribution's tails with fewer realizations. This approach involves summing up the decay times until a given time, resulting in a smoother curve less influenced by fluctuations in individual decay times. 
\begin{figure}
  \centering 
\includegraphics[width=0.99\linewidth]{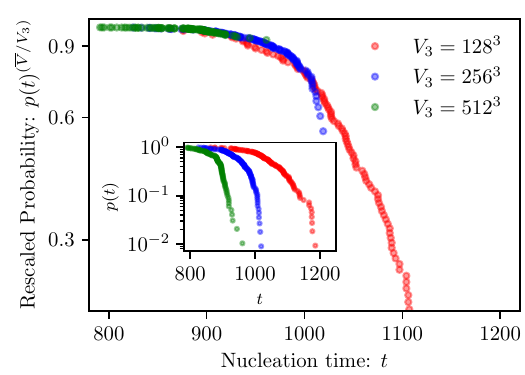}
\caption{Survival probability $p(t)$ rescaled by volume ratio $\overline{V}/V_3$, with $\overline{V}=128^3$, plotted on a log-linear scale for several volumes $V_3 = 128^3, 256^3, 512^3$. The inset shows the same curves without rescaling on a log-linear scale.}
\label{Cumulative_distribution}
\end{figure}
In Fig.~\ref{Cumulative_distribution}, we present the rescaled survival probability $p(t)^{\overline{V}/V_3}$ on a log-linear scale as a function of time for various volumes $V_3$.
The rescaling of the distributions is obtained using a reference volume $\overline{V} = 128^3$. The inset of the figure displays the same functions before the rescaling. Notably, noticeable variations can observed across different volumes. In larger volumes, bubble nucleation occurs earlier, and time distributions are narrower. This observation aligns with the intuition that larger systems are more likely to nucleate bubbles sooner. In the hypothetical scenario of an infinitely large lattice volume, the distribution would resemble a step function akin to a theta function.

Importantly, Fig.~\ref{Cumulative_distribution} demonstrates that, after the rescaling, all curves lie approximately on top of each other. This is consistent with the prediction that the decay rate per unit volume must be volume independent.

Using a range of grid sizes, we could observe a broader range of times compared to what would have been possible with a single grid. This combined distribution allows for a collective analysis. 
Since a single decay rate is not feasible for the considered choice of the parameters, we instead consider a time-dependent $\Gamma(t)$ and extract it using 
\begin{equation}
\frac{dp(t)}{dt} = - \Gamma(t) p(t) \, .
\label{Survival}
\end{equation}
The outcome of this analysis, i.e., the corresponding decay rate $(\Gamma(t)/V_3)^{1/4}$, is presented in Fig.~\ref{fig:methodscomparison}, where the decay rates for $V_3=128^3, 256^3, 512^3$ are shown in the same colors as in Fig.~\ref{Cumulative_distribution}. The standard deviation estimates the error.
\subsubsection{Field expectation value}

\begin{figure}[t]
\centering
\includegraphics[width=0.99\linewidth]{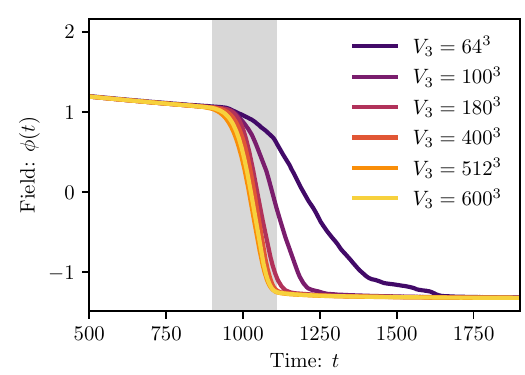}
\caption{One-point function $\phi(t)$ as a function of time $t$ for different lattice volumes $V_3$, ranging from $64^3$ to $600^3$. In the large volume limit, the one-point function becomes volume independent. }
\label{Volume_dependence_opf_prova.pdf}
\end{figure}

The second method for obtaining the decay rate involves studying the evolution of the one-point function during the transition from one minimum to the other.
Fig.~\ref{Volume_dependence_opf_prova.pdf} illustrates the behavior of the one-point function $\phi(t)$ for various volumes $V_3$. It displays the expected volume dependence for small volumes and volume independence for larger volumes, specifically when $V_3 \geq 400^3$. The gray region indicates the transition time range for the large volume limit. 

We define the time-dependent ratio of expectation values as
\begin{equation}
\label{one_point_function_p}
    p_{\phi}(t) \equiv \frac{\phi(t) - \phi_-}{\phi_+ - \phi_-} \, .
\end{equation}
In this context, the one-point function $\phi$ and the minima $\phi_+$ and $\phi_-$ are all time-dependent quantities. This fact arises from the system settling into the local minimum of the potential $\phi_+$ after an initial transient early time regime. Over time, the potential itself evolves, gradually shifting the minimum $\phi_+$ towards smaller values, as observed in the previous section.
Likewise, the global minimum $\phi_-$ changes over time as all fluctuations must settle and thermalize after the transition.
To determine the values of $\phi_+$ and $\phi_-$, we employ polynomial fitting of the one-point function. We fit over a selected time range during which the field settles in the respective minimum of the potential, forming nearly a flat line in the field.
By extrapolating these fits to the entire time range after the early time regime, not relevant to our analysis, we accurately determine the values of $\phi_+$ and $\phi_-$.
This approach ensures that $p_{\phi}(t) $ takes the value of 1 in the false vacuum and 0 in the true vacuum, as expected for a probability function.
The probability $p_{\phi}(t)$ can be expressed as (for a detailed derivation, see~\cite{PhysRevD.23.876, Laine:2016hma})
\begin{align}
p_{\phi}(t) = e^{ - I(t)},
\end{align}
where, in the most general form, the function $I(t)$ is given by
\begin{equation}
I(t)=\frac{4 \pi}{3} \int_{t_0}^t d t^{\prime} \frac{\Gamma_{\phi}(t^{\prime})}{V_3}r^3\left(t, t^{\prime}\right) \, .
\label{eq:volume fraction}
\end{equation}
In this expression, $t_0$ indicates the starting time of the transition. The rate is multiplied by the volume of the bubble of radius $r(t,t')$ formed at time $t'$. The radius at time $t$ of a bubble nucleated at time $t'$ with velocity $v_w$ is approximately given by $r(t,t') \approx v_w(t-t')$, neglecting the bubble size at nucleation.
Note that the previous analysis focuses exclusively on bubble nucleation and expansion while ignoring bubble collisions.

Finally, from \labelcref{eq:volume fraction}, one can compute the decay rate per unit volume as
\begin{equation}
\frac{\Gamma_{\phi}(t)}{V_3} = (8 \pi)^{-1} I^{(4)}(t) \, ,
\end{equation}
where $I^{(4)}$ is the fourth derivative of the function $I(t)$, and we set $v_w =1$ for simplicity. 
Remarkably, this result is independent of the particular choice for $t_0$.

\begin{figure}[t]
    \centering     \includegraphics[width=0.99\linewidth]{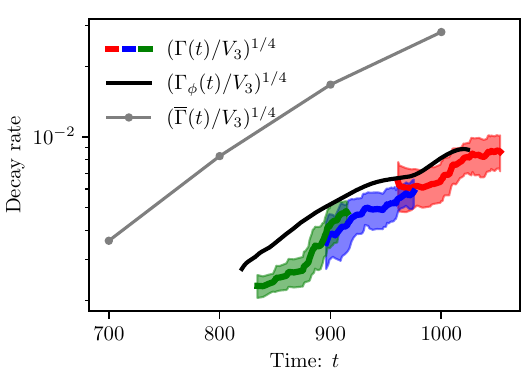}
    \caption{Decay rate $(\Gamma(t)/V_3)^{1/4}$ (from the survival probability), $(\Gamma_{\phi}(t)/V_3)^{1/4}$ (from the evolution of the one-point function) and $(\overline{\Gamma}(t)/V_3)^{1/4}$ (from the bounce) as a function of time. }
    \label{fig:methodscomparison}
\end{figure}

The resulting curve for the corresponding decay rate $(\Gamma_{\phi}(t)/V_3)^{1/4}$ is shown in Fig.~\ref{fig:methodscomparison} as the black solid curve.
We display the decay rate over a period of time starting when the rate starts to be non-negligible and ending when about half of the volume shifts to the new phase, around $p_{\phi} \approx 0.5$.

Before we conclude, we can formulate simple estimates to understand the order of magnitude of the rate. At this stage, we neglect the time dependence of $\Gamma_{\phi}$ for simplicity.
Let us consider the transition time denoted by $\Delta t \equiv t_f- t_0$, i.e., the time required to transition from the start (at $t_0$) to completion (at $t_f$). In that case, the general expression \labelcref{eq:volume fraction} simplifies to
\begin{equation}
	  I(t)  = \frac{\pi}{3} \frac{\Gamma_\phi}{V_3}  v_w^3 (t-t_0)^4 \, ,
\end{equation}
and therefore, for $t = t_f, I(t_f) \approx 1$, leading to 
\begin{equation}
    1  = \frac{\pi}{3} \frac{\Gamma_\phi}{V_3}  v_w^3 \Delta t^4 \, ,
\end{equation}
which gives us the transition time estimate in terms of the rate per unit volume as
\begin{equation}
\label{transition_time}
   \Delta t =  \Bigg( \frac{\Gamma_\phi}{V_3} \frac{\pi}{3} v_w^3 \Bigg )^{-1/4} \approx \Bigg(\frac{\Gamma_{\phi}}{V_3} \Bigg)^{-1/4} \, .
\end{equation}

In the case that the decay proceeds by a single bubble nucleating and expanding, $I$ simplifies to
\begin{equation}
    I(t) \approx \Gamma_\phi (t-t_0) \, ,
\end{equation}
which gives
\begin{equation}
	\Delta t \approx \Gamma_\phi^{-1} \, .
\end{equation}
Therefore, the inverse timescale of the transition, i.e., the rate $\Gamma_{\phi}=  \Delta t^{-1}$, can be estimated as $\Gamma_{\phi} \approx p^{\prime}_{\phi}(t)/p_{\phi}(t)$. 

\subsection{Comparison}

Now that we have all the necessary elements, we can compare real-time decay rates and those computed using the instanton method.

Fig.~\ref{fig:methodscomparison} summarizes the decay rates obtained using the three methods. 

\begin{figure*}[t]
	\centering
	\begin{minipage}[t]{0.49\textwidth}
		\centering
		\includegraphics[width=\linewidth]{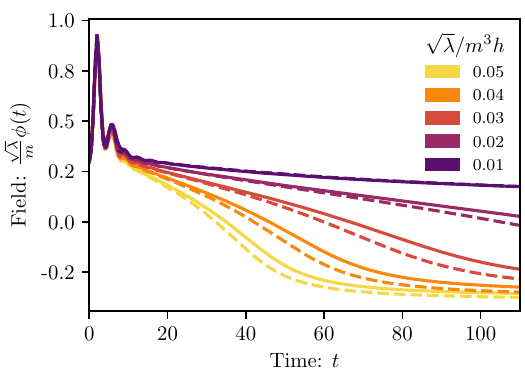}
	
		\label{fig:2PF CS}
	\end{minipage}	~
	\begin{minipage}[t]{0.49\textwidth}
		\centering
		\includegraphics[width=\linewidth]{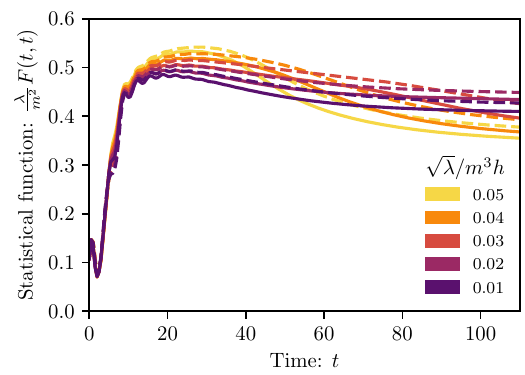}
	
		\label{eq:label}
	\end{minipage}
	\caption{Transition from a false vacuum computed using 2PI NLO (dashed lines) and its approximation to the classical-statistical dynamics (solid lines). \textbf{Left:} One-point function $\phi(t)$. \textbf{Right:} Equal-time statistical two-point function $F(t,t)$. The simulations were realized for different initial conditions for the parameter controlling the asymmetry of the potential $h$. We employ $m^2 = -4, \lambda=30, \frac{\sqrt{\lambda}}{m}\phi_0 = 0.3$.}
	\label{fig:oneandtwoPF_2PI}
\end{figure*}

Using the survival probability method, we can combine the results obtained using different volumes. Remarkably, within the overlapping regimes, the corresponding rate $(\Gamma(t)/V_3)^{1/4}$ is volume independent and nicely forms a single line.

Furthermore, the figure exhibits an intriguing trend, common to all methods, showing that the decay rates increase exponentially in time. 
Both real-time methods, the survival probability one and the one-point function method, show good quantitative agreement. In contrast, the instanton method exhibits a slight shift from the other two curves, possibly attributed to the neglected pre exponential factors. 
Including this prefactor was beyond the scope of our analysis. Nevertheless, the prefactor is expected to be less sensitive to the time-dependent shape of the potential compared to the exponential term. This is indeed reflected in the fact that both the instanton and the real-time methods predict a similar time evolution of the decay rate, and the quantitative agreement of the decay rate is quite satisfactory. 

We verified that when using the thin wall approximation of the bounce, the agreement between the two methods becomes worse. This is reflected in both the magnitude of the decay rate as well as its time dependence.
\section{Towards the quantum regime of real-time false vacuum decay}
\label{sec:Quantum}

In the previous section, we focused on investigating a situation where the decay of the false vacuum takes place at high temperatures. This scenario allows for the application of classical-statistical simulations to describe the dynamics. In this context, the decay was observed to occur through bubble nucleation, and we analyzed real-time observables. 

In this section, our aim is to formulate the problem of false vacuum decay directly in quantum field theory. The description is based on the nonequilibrium 2PI effective action on a closed time path. It describes the time evolution in terms of dynamical equations for one- and two-point correlation functions of the field.

\subsection{Quantum effective action}

The 2PI effective action, denoted by $\Gamma[\phi, G]$, is formulated in terms of a one-point function, $\phi(x)$, and a two-point function, $G(x,y)$. It can be expressed as follows~\cite{berges2015nonequilibrium}:
\begin{equation}
	\begin{aligned}
		\Gamma[\phi, G] =&  \,\,\,  S_\mathcal{C}[\phi]
  +\frac{i}{2} \operatorname{Tr}_{\mathcal{C}} \ln [G^{-1}]
		+  \frac{i}{2} \operatorname{Tr}_\mathcal{C}[G_0^{-1}(\phi) G] \\ & +\Gamma_2[\phi,G]+\text {const\,,}
	\end{aligned}
	\label{Quantumeom}
\end{equation}
where $S_C[\phi]$ is the classical action integrated over the Schwinger-Keldysh contour, $G_0^{-1}(\phi)$ is the inverse classical propagator, and $\Gamma_2[\phi, G]$ is the sum of all 2PI graphs with full propagator lines. 

The quantum equations of motion can be derived by taking the functional derivative of the effective action $\Gamma[\phi, G]$ with respect to the macroscopic field $\phi(x)$ and the propagators $G(x,y)$. The stationary conditions are then obtained by setting these functional derivatives to zero, resulting in a system of coupled equations.

The equations can be expressed by decomposing the two-point function into its statistical component $F(x,y)$ and spectral component $\rho(x,y)$ as \footnote{
	The spectral propagator is defined as $\rho(x, y)=i\left\langle\left[\varphi(x), \varphi(y)\right]\right\rangle $ and is related to the possible states. In contrast, the statistical propagator is defined as $F(x, y)=\frac{1}{2}\left\langle\left\{\varphi(x),\varphi(y)\right\}\right\rangle - \phi(x)\phi(y)$ and encodes information relative to the occupancies.}~\cite{Aarts2001NonequilibriumTE}
\begin{equation}
	G(x, y)=F(x, y)+\frac{i}{2} \rho(x, y) \operatorname{sgn}_\mathcal{C}\left(x^0-y^0\right) \, ,
 \label{eq:2PFdecomposition}
\end{equation}
and are given by
\begin{equation}
	\begin{aligned}
		{\left[\square_x +M^2(x)\right] F(x, y)=} &-\int_{t_0}^{x^0} \mathrm{~d} z \Sigma^\rho(x, z) F(z, y) \\
		&+\int_{0}^{y^0} \mathrm{~d} z \Sigma^F(x, z) \rho(z, y) \, , \\
		{\left[\square_x +M^2(x)\right] \rho(x, y)=} &-\int_{y^0}^{x^0} \mathrm{~d} z \Sigma^\rho(x, z) \rho(z, y) \, ,
	\end{aligned}
	\label{2PIeq}
\end{equation}
 and
\begin{equation}
	\begin{aligned}
\left[\square_x+M^2(x)+h\right] \phi(x)=- \int_{0}^{x^0} \mathrm{~d} z \Sigma^{\rho}(x, z; \phi=0) \phi(z) \, ,
	\end{aligned}
	\label{eq:2PIeqfield}
\end{equation}
where we indicate $\int_{t_1}^{t_2} d z=\int_{t_1}^{t_2} d z^0 \int d^3 z$. The effective mass squared term $M^2(x)$ in the above equations represents a correction from the presence of the background field and the fluctuations and is given by
\begin{equation}
    \begin{aligned}
        M^2(x) & =m^2 +\Sigma^{(0)}(x) \, .
    \end{aligned}
\end{equation} 
The self energies $\Sigma^{\rho}$, $\Sigma^F$, and $\Sigma^{(0)}$ are defined as follows:
\begin{equation}
    \begin{aligned}
        2 i \frac{\delta \Gamma_2[\phi, G]}{\delta G(x, y)} = &-i \Sigma^{(0)}(x) \delta(x-y)+\Sigma^F(x, y) \\ & -\frac{i}{2} \Sigma^\rho(x, y) \operatorname{sgn}_{\mathcal{C}}\left(x^0-y^0\right) \, ,
    \end{aligned}
\end{equation}
where the dependence of the self energies on $\phi$ and $G$ is implied. The right-hand sides of the evolution equations involve memory integrals, which take into account the whole time history of the evolution.

To solve the set of coupled equations \labelcref{2PIeq} and \labelcref{eq:2PIeqfield}, the self-energies must be truncated. A powerful nonperturbative truncation scheme is provided by the 2PI large-$N$ expansion~\cite{berges_2002,aarts_2002}. Already the NLO of this expansion is known to be able to describe various phenomena observed for nonequilibrium scalar fields, such as instabilities~\cite{Berges:2002cz, Arrizabalaga:2004iw} and thermalization~\cite{Berges:2001fi, Juchem:2003bi}. Even though the convergence of this expansion relies on a large value of $N$, the method has been successfully applied to models with a moderate number of field components and even for single-component fields.

In the following subsection, we present some numerical results for the dynamics of the false vacuum decay within the NLO large-$N$ expansion, focusing on a single field component. The explicit form of the equations of motion is presented in Appendix~\ref{app:1/Nexp}, along with the corresponding expressions for the self-energies within the chosen approximation scheme. To fully determine the system's time evolution described by the equations of motion, the initial conditions must be specified, as outlined Appendix~\ref{app:1/Nexp} [\labelcref{eq:2PIinitialconditions1PF} and \labelcref{eq:2PIinitialconditions2PF}].

\subsection{2PI NLO evolution for vacuum initial conditions} \label{Quantum:CSA}

In this section, we consider general nonequilibrium initial conditions where—unlike before—the field does not reach a prethermalized state before the transition. Therefore, it is not possible to estimate the decay rate per unit volume using the same assumptions as in the previous analysis. 

In such circumstances, we focus on the dynamics of field correlation functions. By studying these correlation functions, we can track the decay of the field from the false minimum without referring to notions such as bubbles, which are well defined only in the semiclassical limit. 

In our simulations we set $m^2 = -4$, $\lambda = 30$, $\sqrt{\lambda}\phi_0 /m = 0.3$, $(1 \times 32)^3$ lattice, and employ vacuum initial conditions ($f_{\boldsymbol{p}}=0$). The sole varying parameter is the linear coupling $h$, which affects the potential's tilt strength. 

A powerful advantage of the 2PI effective action approach is that it allows a clear comparison between the quantum dynamics and the classical-statistical approximation. In order to detect both qualitative and quantitative indications of quantum effects in the dynamics, we run the simulations twice with identical initial conditions within the NLO large-$N$ approximation. First, by simulating the entire quantum dynamics and, second, only the statistical-classical limit of the dynamics (for further explanation, see Appendix~\ref{app:CSapproximation}). The results are shown in Fig.~\ref{fig:oneandtwoPF_2PI}, where the dynamics of the one- and two-point functions is plotted. The full quantum dynamics (classical-statistical limit of the dynamics) is indicated by dashed lines (solid lines).
As can be observed, the field's decay in the full quantum dynamics occurs faster than in the corresponding classical-statistical limit.
This observation suggests that the quantum part in the dynamics, although challenging to isolate due to the complexity of the equations, speeds up the transition.

\begin{figure}[t]
\centering
\includegraphics[width=0.99\linewidth]{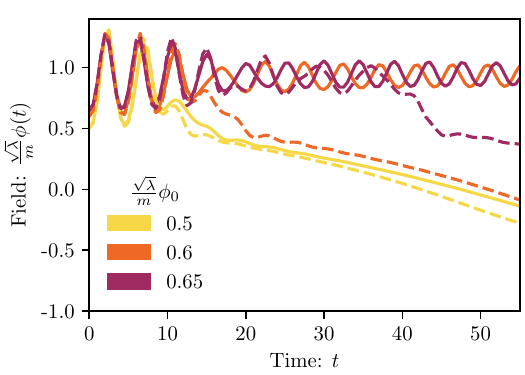}
\caption{Time evolution of the one-point function $\phi(t)$ using 2PI $1/N$ to NLO for the full quantum evolution (dashed lines) and the classical-statistical approximation (solid lines). The simulations are performed for various initial field values $\phi_0$. We employ $m^2 = -4, \lambda = 10, h = 0.01$. }
\label{FQvsCS1N}
\end{figure}

Next, we explore a variation in the initial displacement of the one-point function from the false minimum. We employ $m^2 = -4, \lambda = 10, h=0.01$, and $ V_3 = (0.5 \times 64)^3$. We keep these parameters fixed and vary the initial value of the macroscopic field $\phi_0$. 
The numerical results are depicted in Fig.~\ref{FQvsCS1N}: the full quantum evolution (the classical-statistical approximation) is indicated by dashed lines (solid lines).
\begin{figure*}[t]
	\centering
	\begin{minipage}[t]{0.49\textwidth}
		\centering
		\includegraphics[width=\linewidth]{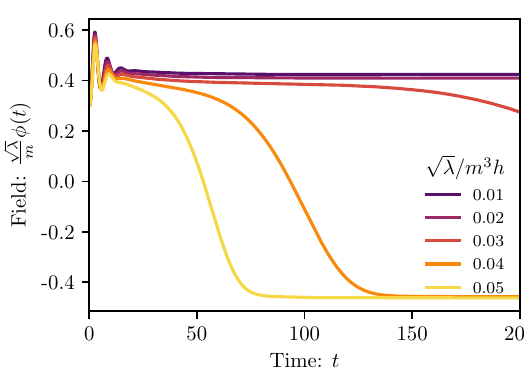}
		\label{fig:G_rescaled}
	\end{minipage}	~
	\begin{minipage}[t]{0.49\textwidth}
		\centering
		\includegraphics[width=\linewidth]{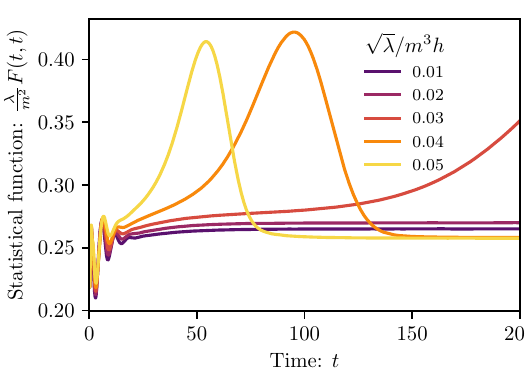}
		\label{eq:label}
	\end{minipage}
	\caption{Transition from a false vacuum in full classical-statistical simulations. \textbf{Left:} Evolution of the one-point function $\phi(t)$ for different linear couplings $h$.  \textbf{Right:} Equal-time statistical two-point function $F(t,t)$ for different linear couplings $h$.  We employ the same parameters as in Fig.~\ref{fig:oneandtwoPF_2PI}, and $V_3=(1 \times 256)^3$, for various $h$ as shown in the legend. Each curve is an ensemble of 20 runs.}
	\label{fig:oneandtwoPF_CSA}
\end{figure*}
We start with an initial condition for the field such that its amplitude generates enough fluctuations that lead to the decay of the field. In this case, the field crosses the barrier almost immediately for very large initial field oscillation amplitudes. Then, we progressively reduce the field's initial amplitude to have fewer fluctuations generated by the oscillations. The field experiences parametric resonance followed by damped oscillations and remains in the metastable state for a longer period of time. 

 In the case of the full NLO evolution, the field oscillates and, eventually, depending on the initial conditions, transitions. In contrast, in the NLO classical-statistical limit, there is a threshold value ($|\phi_0| > 0.5 $) beyond which the system does not decay at all in our simulations but continues oscillating in the well.\footnote{The decay in the NLO classical-statistical dynamics could happen in principle much later, but this was not observed in the simulated timescales.} Therefore, in this case, we can say that quantum effects aid in the decay process.

Based on these observations, we can conclude that even if the system does not possess sufficient energy to surpass the potential barrier in the presence of classical-statistical fluctuations, quantum fluctuations drive the transition to the other side. A similar behavior was observed also in the case of initial conditions with a small displacement of the one-point function from the minimum but, instead, stronger fluctuations encoded in the two-point function. 

\subsection{Assessing NLO $N=1$ approximation accuracy} \label{subsec:comparison}

Having compared the classical-statistical and the quantum dynamics of the false vacuum decay within the NLO approximation, we now want to check the accuracy of this approximation for $N=1$.
To do so, we simulate the full classical-statistical dynamics using the same parameters as in Fig.~\ref{fig:oneandtwoPF_2PI} ($m^2 = -4, \lambda=30, \sqrt{\lambda}\phi_0/m = 0.3, f_{\boldsymbol{p}}=0$).  The lattice volume was chosen large enough such that the results are volume independent. 
 
The results from the classical-statistical simulations are illustrated in Fig.~\ref{fig:oneandtwoPF_CSA} by solid lines. The overall behavior of the dynamics of the one-point function is qualitatively the same as in the previous cases, with the NLO classical-statistical usually showing a faster evolution than the full classical statistical approach for this choice of parameters. 
The two-point function also shows different behavior: in full classical-statistical simulations, the graph shows a clearer peak during the field decay, which is not observed in NLO classical-statistical simulations. 

One reason for this discrepancy of the large-$N$ expansion for $N=1$ is that higher-order contributions such as NNLO are no longer suppressed. 
We investigate the discrepancy between the two methods further in Appendix~\ref{app:boxIC}. There we present a comparison between classical-statistical and NLO classical-statistical time evolution of the one-point function for a different choice of initial conditions for the fluctuations in terms of box initial conditions.

\section{Conclusions and outlook} \label{sec:Conclusion}

In our study, we presented a real-time approach to investigate the phenomenon of false vacuum decay in quantum field theory. We focused on the dynamics of a relativistic scalar field in $3+1$ dimensions, initialized near a local minimum of a tilted double-well potential. 

First, we considered a scenario where strong fluctuations are present. We employed the classical-statistical approximation to simulate the dynamics of the system in this regime. By this, we demonstrated that the real-time decay rates are comparable to those obtained from the bounce solution of the conventional imaginary-time approach. In general, we found that the decay rates are time dependent. Correspondingly, we extracted a time-dependent effective potential, which becomes convex during the nonequilibrium transition process.

Second, we extended our analysis to the case of strongly coupled systems with low occupancy, where quantum effects are expected to play an important role. We used the nonequilibrium quantum evolution equations derived from the irreducible effective two-particle action truncated at NLO in $1/N$ to study the dynamics of the field and two-point functions. We demonstrated that quantum corrections can lead to transitions that are not captured by classical-statistical approximations.

Our results help bridging the nonequilibrium quantum field theory and the Euclidean approach to false vacuum decay. In contrast to the latter, the real-time method allows one to address the initial-value problem with the different stages of the subsequent transition. False vacuum decay is often not captured in terms of a (constant) decay rate but may be more efficiently characterized in terms of a time-dependent effective potential, which describes the transition from the metastable phase to the stable minimum. The notion of a time-dependent effective potential for a quantum description of false vacuum decay should be accessible in controlled experiments with ultracold quantum gases by measuring higher-order correlation functions~\cite{Zache:2019xkx, Prufer:2019kak}.

\subsection*{ACKNOWLEDGMENTS}
We thank Jonathan Braden, Sebastian Erne, Stefan Flörchinger, Michael Heinrich, Louis Jussios and Jörg Schmiedmayer for valuable discussions. 
The authors acknowledge support by the state of Baden-Württemberg through bwHPC and the German Research Foundation (DFG) through Grant No. INST 40/575-1FUGG (JUSTUS 2 cluster). 
L.B .and J.B. acknowledge support by the DFG under the Collaborative Research Center SFB
1225 ISOQUANT (Project-ID No. 27381115) and the Heidelberg STRUCTURES Excellence Cluster under Germany’s Excellence Strategy EXC2181/1-390900948.
A.C. acknowledges support by the Deutsche Forschungsgemeinschaft under Germany’s Excellence Strategy – EXC 2121, Quantum Universe – 390833306. Nordita is supported in part by NordForsk.
\appendix  \label{Appendix}

\section{Nonequilibrium quantum field theory} \label{NEQQFT}
To describe the time evolution of the system starting from an initial nonequilibrium state, we introduce a functional known as the generating functional, denoted as $Z[J, R;\varrho(t_0)]$~\cite{Berges2005NonequilibriumQF}. Here, $\varrho(t_0)$ represents the system's density matrix at the initial time $t_0$, while $J$ and $R$ are external sources coupled to the field operator $\varphi$. The generating functional is defined as follows:

\begin{equation}
\begin{aligned}
Z[J,R;\varrho(t_0)]  = & \,\,\, \operatorname{Tr} \Big\{ \varrho(t_0)T_{\mathcal{C}} e^{i \int_{x, \mathcal{C}} J(x)\varphi(x)} \\ &
\times e^{\frac{i}{2}\int_{x,y, \mathcal{C}} \varphi(x) R(x,y) \varphi(y)} \Big\} \, .
\label{eq:pathintegral}
\end{aligned}
\end{equation}
The time integral is performed along the closed time path $\mathcal{C}$. It starts at $t_0$, continues forward on the real-time axis to $t_{f}$ (the forward branch $\mathcal{C}_+$) and then moves back on the real-time axis to $t_0$ (the backward branch $\mathcal{C}_-$). The time-ordering operator $T_{\mathcal{C}}$ ensures that the operators are correctly ordered along the contour with respect to time direction.

In the case of Gaussian initial states, the dependence on the density matrix $\varrho(t_0)$ is fully accounted for by providing initial conditions for $\phi$ and $G$.

Moreover, the closed time contour $\mathcal{C}$ can be separated into the forward and backward time integrations, and the fields $\varphi(x)$ can be expressed as a sum of the forward and backward fields $\varphi^+(x)$ and $\varphi^-(x)$, respectively, which are taken on the $\mathcal{C}^+$ and $\mathcal{C}^-$ branches of the contour. This allows us to rearrange the fields in terms of these linear combinations
\begin{equation}
	\varphi(x)=\frac{\varphi^{+}(x)+\varphi^{-}(x)}{2} \,, \quad \widetilde{\varphi}(x)=\varphi^{+}(x)-\varphi^{-}(x) \, .
\end{equation}

All time-ordered connected correlation functions can be obtained by taking functional derivatives of the Schwinger functional $W[J, R]$ defined as
$
Z[J, R]=e^{i W[J, R] } \, ,
$ 
with respect to the source terms and then setting the sources to zero. Specifically, the one-point function, or macroscopic field, reads
\begin{equation}
	\phi(x) \equiv \langle  \varphi(x) \rangle = \frac{\delta W[J,R]}{\delta J(x)}\Bigg|_{J, R=0},
	\label{eq:phiaverage}
\end{equation}
and the connected two-point function, or propagator,  
\begin{equation}
	\begin{aligned}
		G(x, y) & \equiv  \left\langle T_\mathcal{C} {\varphi}(x) {\varphi}(y)\right\rangle-\phi(x) \phi(y) \\  & 
		=\left.2 \frac{\delta W[J, R]}{\delta R(x, y)}\right|_{J, R=0}-\phi(x) \phi(y) \, .
	\end{aligned}
\end{equation}
	\section{ Evolution equations } \label{app:1/Nexp}
	
  The equations presented in Sec.~\ref{sec:Quantum} are derived from first principles and capture the whole quantum evolution. However, due to their complexity, they generally cannot be solved without resorting to approximations. Therefore, we introduce a systematic nonperturbative approximation scheme known as the $1/N$ expansion. This appendix presents the evolution equations derived from the 2PI effective action under the $1/N$ expansion, truncated at NLO. See, e.g, ~\cite{Shen:2020jya} for more details.

To illustrate different approximation schemes in subsequent sections, we will consider an example of a quantum field theory involving a real, $N$-component scalar field $\varphi_a(x)$ with an $O(N)$-symmetric classical action
\begin{equation}
	\begin{aligned}
		S_\mathcal{C}[\varphi]  =\int_x &\Big\{\frac{1}{2} \partial^\mu \varphi_a(x) \partial_\mu \varphi_a(x)+\frac{m^2}{2}  \varphi_a(x) \varphi_a(x) \\& -\frac{\lambda}{4 ! N}\left(\varphi_a(x) \varphi_a(x)\right)^2\Big\} \, .
	\end{aligned}
\end{equation}
corresponding to a potential as in \labelcref{Bare} for a single component field. We specifically concentrate on the case of $N=1$ from now on.

The explicit form of the self energies in \labelcref{2PIeq} are given by 
\begin{equation}
     \Sigma^{(0)}(x) =  \frac{\lambda}{2}\Big(F(x, x)+\phi^2(x) \Big) \, ,
     \label{eq:B2}
\end{equation}
\begin{equation}
    \begin{aligned}
\Sigma^F(x, y) = &-\frac{\lambda}{3 }\Bigg\{I_F(x, y)\left[\phi(x) \phi(y)+F(x, y)\right] \\ 
& \left.-\frac{1}{4} I_\rho(x, y) \rho(x, y) +P_F(x, y) F(x, y)\right. \\ &
-\frac{1}{4} P_\rho(x, y) \rho(x, y)\Bigg\} \, ,
\end{aligned}
\end{equation}
and
\begin{equation}
    \begin{aligned}
\Sigma^\rho(x, y) = & -\frac{\lambda}{3 } 
 \Bigg\{ I_\rho(x, y)\left[\phi(x) \phi(y)+F(x, y)\right]
\\ & +I_F(x, y) \rho(x, y) +P_\rho(x, y) F(x, y) \\ &
+P_F(x, y) \rho(x, y) \Bigg\} \, .
   \end{aligned}
\end{equation}

In the previous equations, we denote the statistical and spectral components of the summation functions as
\begin{equation}
\begin{aligned}
&I_F(x, y)  =  \frac{\lambda}{6}\left(F^2(x, y)-\frac{1}{4} \rho^2(x, y)\right) \\
 & - \frac{\lambda}{6}\Bigg \{\int_0^{x^0} \hspace{-.7em} 
 \mathrm{~d} z \, I_\rho(x, z)\left(F^2(z, y)-\frac{1}{4} \rho^2(z, y)\right) \\ & -2 \int_0^{y^0} \hspace{-.7em} 
 \mathrm{~d} z \, I_F(x, z) F(z, y) \rho(z, y)\Bigg\} \, ,
\end{aligned}
\end{equation}
where we set $t_0 = 0$, and 
\begin{equation}
\begin{aligned}
I_\rho(x, y) =& \,\,\, \frac{\lambda}{3} F(x, y) \rho(x, y) \\ &
-\frac{\lambda}{3} \int_{y^0}^{x^0} \hspace{-.7em} \mathrm{~d} z  \,
 I_\rho(x, z) F(z, y) \rho(z, y) \, ,
\end{aligned}
\end{equation}
with the field dependent resummation functions given by
\begin{equation}
 \begin{aligned}
    & P_F(x, y)  =-\frac{\lambda}{3 }\Bigg\{H_F(x, y) \\& 
     -\int_0^{x^0} \hspace{-.7em}  \mathrm{~d} z\left[H_\rho(x, z) I_F(z, y) 
     +I_\rho(x, z) H_F(z, y)\right]  \\
& +\int_0^{y^0} \hspace{-.7em} 
 \mathrm{~d} z\left[H_F(x, z) I_\rho(z, y) +I_F(x, z) H_\rho(z, y)\right]
\\ &-\int_0^{x^0} \hspace{-.7em} 
 \mathrm{~d} z \int_0^{y^0} \hspace{-.7em} \mathrm{~d} v I_\rho(x, z) H_F(z, v) I_\rho(v, y) \\
& +\int_0^{x^0} \hspace{-.7em} 
 \mathrm{~d} z \int_0^{z^0} \hspace{-.7em}  \mathrm{~d} v \, I_\rho(x, z) H_\rho(z, v) I_F(v, y) 
\\
& +\int_0^{y^0} \hspace{-.7em} 
 \mathrm{~d} z \int_{z^0}^{y^0} \hspace{-.7em}  \mathrm{~d} v \, I_F(x, z) H_\rho(z, v) I_\rho(v, y)\Bigg\} \, ,
\end{aligned}
 \end{equation}
 \begin{equation}
     \begin{aligned}
P_\rho(x, y)= & -\frac{\lambda}{3 }\Bigg\{H_\rho(x, y)\\ & -\int_{y^0}^{x^0} \hspace{-.7em} 
 \mathrm{~d} z \, \left[H_\rho(x, z) I_\rho(z, y) +I_\rho(x, z) H_\rho(z, y)\right]
 \\ & +\int_{y^0}^{x^0} \hspace{-.7em}  \mathrm{~d} z \int_{y^0}^{z^0} \hspace{-.7em} 
 \mathrm{~d} v \,\, I_\rho(x, z) H_\rho(z, v) I_\rho(v, y)\Bigg\} \, ,
\end{aligned}
 \end{equation}
and the field dependence is contained in the functions 
 \begin{align}
     H_F(x, y) \equiv-\phi(x) F(x, y) \phi(y) \, , \\
     H_\rho(x, y) \equiv-\phi(x) \rho(x, y) \phi(y) \, .
     \label{eq:B10}
 \end{align}

For practical computations, we assume spatial isotropy and homogeneity. Therefore, the integrals in momentum space will only depend on a single radial momentum.

Initial conditions must be prescribed to solve \labelcref{2PIeq} and \labelcref{eq:2PIeqfield}.
The initial conditions for the macroscopic field can be written as
	\begin{equation}
		\phi(t)|_{t=0}= \phi_0 \,, \quad \partial_t \phi(t)|_{t=0}=0 \, .
    \label{eq:2PIinitialconditions1PF}
	\end{equation}

 \begin{figure*}[t]
\includegraphics[width=.49\linewidth]{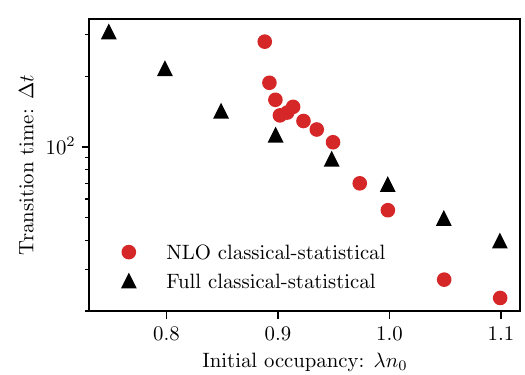}
\includegraphics[width=.49\linewidth]{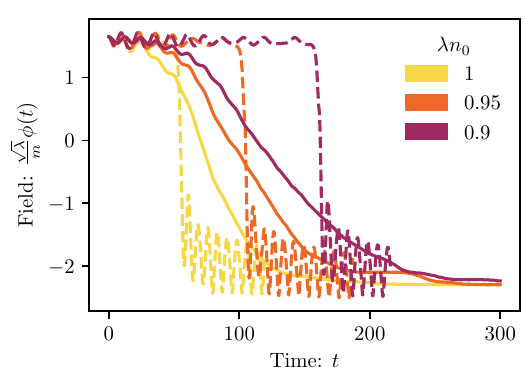}
    \caption{Comparison of the full classical-statistical with NLO classical-statistical simulations for box initial conditions \labelcref{eq:boxICoccupations}. \textbf{Left:} Transition timescale $\Delta t$ in terms of the initial occupancy $n_0$ for NLO classical-statistical (red dots) and full classical-statistical simulations (black triangles) obtained averaging over 100 runs. The NLO classical-statistical simulations indicate around $\lambda n_0  \approx 0.9$  an asymptotic value where the transition timescale diverges. \textbf{Right:} Evolution of the one-point function $\phi(t)$ for three selected initial conditions shown in the left panel. The dashed lines indicate the full classical-statistical and NLO classical-statistical are indicated by solid lines. We employ $h = 0.9, \frac{\sqrt{\lambda}}{m}\phi_0 = 1.65$. }
    \label{fig:boxIC}
\end{figure*}

 Gaussian initial conditions for the statistical two-point function considered in this work correspond to	
\begin{equation}
\begin{aligned}
    &F(t,t', |\boldsymbol{p}|) |_{t = t'=0}  =\frac{f_{\boldsymbol{p}}+1/2}{\sqrt{|\boldsymbol{p}|^2+M^2}} \, , \\
    &\partial_t F(t, t'; |\boldsymbol{p}|)_{t=t'=0}  =0 \, , \\
   &  F(t,t', |\boldsymbol{p}|) \partial_t \partial_{t^{\prime}} F\left(t, t^{\prime}, |\boldsymbol{p}|\right)|_{t=t^{\prime}=0}  =\left[f_{\boldsymbol{p}}+1/2 \right]^2 \, .
   \label{eq:2PIinitialconditions2PF}
\end{aligned}
\end{equation}
 In Sec.~\ref{sec:Quantum}, we have chosen the initial quasiparticle occupation in vacuum, i.e., $f_{\boldsymbol{p}} =0$. 
 The initial conditions for the spectral function are determined by its antisymmetry, $\rho(t, t,|\boldsymbol{p}| ) = 0$, and the equal-time commutation relations of the underlying bosonic quantum theory.
 
 The source code used for the 2PI simulations is accessible in the Zenodo repository under~\cite{shen_linda_2020_3698136}.
The numerical time evolution is computed using a spatial grid of $N_x=64$
lattice points and a lattice spacing of $a_x = 0.5$, unless stated differently. Moreover, the time step is chosen to be $a_t = 0.01 a_x$, guaranteeing energy conservation at the level of a few percent.

	\section{Classicality condition} \label{app:CSapproximation}
This appendix provides additional details on the classical-statistical field theory framework. 
 In this approach, the field is treated classically, and the dynamics involves sampling over initial conditions evolving each realization in real time according to the equation of motion of the classical field. Observables are obtained by averaging over classical trajectories. 
 
 A generating functional analogous to the one discussed in \labelcref{eq:pathintegral} can be formulated for computing correlation functions within classical-statistical field theory. This generating functional enables us to calculate various statistical quantities and correlations.
For a more comprehensive and detailed explanation of the subject, see~\cite{berges2015nonequilibrium}.
 
The exact time evolution equations for $\rho_{\mathrm{cl}}$ and $F_{\mathrm{cl}}$ share the same structure as in~\labelcref{2PIeq}, featuring self-energies $\Sigma^{\rho}_{\mathrm{cl}}, \Sigma^F_{\mathrm{cl}}$. 

In particular, we are interested in seeing how the previously discussed full 2PI equations of motion differ from their classical-statistical approximation. 
 It is important to consider how the equations governing the evolution of the system are affected in this limit and the effects on the dynamics, which is explored in Sec.~\ref{sec:Quantum}.
One finds that the classical self-energies are obtained from the respective quantum ones as~\cite{berges2015nonequilibrium}
\begin{equation}
\Sigma^F_{\mathrm{cl}}=\Sigma^F\left(F^2 \gg \rho^2\right) \, , \quad \Sigma^{\rho}_{ \mathrm{cl}}=\Sigma^\rho\left(F^2 \gg \rho^2\right) \, .
\end{equation}
The classical-statistical self-energies are derived from their quantum counterparts by neglecting terms involving the product of two spectral functions $\rho$ compared to the product of two statistical propagators $F$. Importantly, this condition does not require thermal equilibrium and can be applied to systems undergoing out-of-equilibrium dynamics, irrespective of the magnitude of the macroscopic field~\cite{Tranberg:2022noe}.

The comparison between quantum and classical-statistical dynamics is made more transparent by rescaling the macroscopic field and statistical correlation function. Specifically, we have the following rescaling
\begin{equation}
    \phi(x) \to \sqrt{\lambda} \phi(x) \, , \quad F(x, y) \to\lambda F(x, y) \, ,
\end{equation}
whereas the spectral function $\rho(x, y)$ remains unchanged. Similarly, the statistical self-energy rescales as $\Sigma^{F}(x, y) \to \lambda \Sigma^F(x, y)$. Because of this rescaling, the classical self-energies also show an important property: they allow us to remove self-interaction influence from the dynamics. As a result, the coupling constant only affects the initial conditions. However, quantum corrections break this property and introduce terms proportional to $\rho^2$, which become more significant as the coupling strength increases.

\section{Starting with box initial conditions} \label{app:boxIC}

This appendix investigates the qualitative and quantitative differences between the full classical-statistical and NLO classical-statistical dynamics in the same spirit of Sec.~\ref{subsec:comparison}. This time, we make a different choice of initial conditions.
In order to skip the parametric resonance stage, it is convenient to start with a fixed nonvanishing background field $\phi_0$ and initial occupation numbers of the form 
\begin{equation}
f_{\boldsymbol{p}}=n_0 \theta(Q-\boldsymbol{p}) \, .
\label{eq:boxICoccupations}
\end{equation}
These initial conditions correspond to a box in momentum space up to the momentum scale $Q=1$, with amplitude $n_0$.
We set $\phi_0 = 1.65 / \sqrt{\lambda}$, which is close to the false minimum of the potential for $h=0.9$. We use $V_3 = (32 \times 1.5)^3$ for the grid. 
In the left panel of Fig.~\ref{fig:boxIC}, we compare the extracted transition timescale $\Delta t$ (that is, the time when the field becomes negative) resulting from the full classical-statistical simulations averaging over 100 runs (black triangles) and the simulations of the $2 \mathrm{PI}$  $1 / N$ NLO classical-statistical (red dots). 
For high initial occupation numbers, the transition time of the full classical-statistical simulations is slower than the one of the NLO classical-statistical simulations. This is due to higher order contribution to the dynamics (NNLO, etc.) as discussed in the main text.
Going towards lower initial occupation numbers, the transition timescales of the NLO classical-statistical simulations become slower than the full classical-statistical case, showing around $\lambda n_0  \approx 0.9$ an asymptotic value where the transition timescale diverges. 
To get more insight into the decay dynamics, we show in the right panel of Fig.~\ref{fig:boxIC} the time evolution of the field for three selected initial conditions from the left panel.
The solid (dashed) lines indicate the full classical-statistical (NLO classical-statistical) field evolution. 
We observe that the full classical-statistical and NLO classical-statistical do not match well. In particular, the NLO classical-statistical dynamics is characterized by a sharp decay and shows oscillations around the true vacuum after the transition. The full classical-statistical dynamics starts decaying earlier, and the transition takes a longer time to occur.
This discrepancy requires further clarification.
\bibliography{references}

\end{document}